\documentclass[twocolumn,showpacs,preprintnumbers,amsmath,amssymb,aps,pre]{revtex4}

\usepackage{mathtools,amssymb,lipsum}

\usepackage{epsfig}

\usepackage{graphicx,latexsym,xcolor}

\usepackage{tikz}

\usepackage[english]{babel}
\usepackage{amsmath}
\usepackage{color}
\usepackage{epsfig}
\usepackage{graphicx}
\usepackage{bm}
\usepackage{times,amsmath,amssymb}
\usepackage{subfigure}
\usepackage{amsfonts}
\usepackage{amssymb}
\usepackage{color}
\global\long\def\bege{\begin{equation}}
\global\long\def\ende{\end{equation}}

\global\long\def\begal{\begin{align}}
\global\long\def\endal{\end{align}}

\begin{document}

\title{Quantum chaos, localization and phase transitions in random graphs}
\author{Ioannis Kleftogiannis$^1$, Ilias Amanatidis$^{2}$}
\affiliation{$^1$ Physics Division, National Center for Theoretical Sciences, Hsinchu 30013, Taiwan }
\affiliation{$^2$Department of Physics, Ben-Gurion University of the Negev, Beer-Sheva 84105, Israel}

\date{\today}
\begin{abstract}
The energy level statistics of uniform random graphs are studied,
by treating the graphs as random tight-binding lattices.
The inherent random geometry of the graphs and their dynamical spatial dimensionality, leads to various quantum chaotic and localized phases and transitions between them. Essentially the random geometry acts as disorder, whose strength is characterized by the ratio of edges over vertices R in the graphs. For dense graphs, with large ratio R, the spacing between successive energy levels follows the Wigner-Dyson distribution, leading to a quantum chaotic behavior and a metallic phase, characterized by level repulsion. For ratios near R=0.5, where a large dominating component in the graph appears, the level spacing follows the Poisson distribution with level crossings and a localized phase for the respective wavefunctions lying on the graph. For intermediate ratios R we observe a phase transition between the quantum chaotic and localized phases characterized by a semi-Poisson distribution. The values R of the critical regime where the phase transition occurs depend on the energy of the system. Our analysis shows that physical systems with random geometry, for example ones with a fluctuating/dynamical spatial dimension, contain novel universal phase transition properties, similar to those occuring in more traditional phase transitions based on symmetry breaking mechanisms, whose universal properties are strongly determined by the dimensionality of the system. 

\end{abstract}

\maketitle

Phase transitions in matter, for example ones that rely on breaking of symmetries, contain characteristics like critical exponents, that are strongly determined by a few generic parameters like the spatial dimension of the system. These so called universalities show that a few parameters like the dimensionality are sufficient to fully determine the phase transition properties without a reference to the microscopic details of the system. From this point of view it is interesting to examine physical systems which do not posses a well defined spatial dimension or systems where the dimension fluctuates or is dynamical, like random graphs/networks\cite{erdos_gallai,aigner,farkas,newman,frieze,berg,mizutaka,paper1,paper2}. Such systems contain random discrete geometries at a microscopic level, that can converge to continuous manifold geometries at macroscopic scales when a lot of vertices(nodes) are considered\cite{paper1,paper2}. We study the phase transition problem by treating the graphs/networks as random tight-binding lattices and then study the statistics of their energy levels by calculating the probability distribution of the spacing between successive energy levels.

In regular disordered systems, for example regular tight-binding lattices with a random on-site potential, such as the Anderson model, destructive interference of the quantum particle waves leads to phenomena like an exponential localization of the wavefunctions. The energy-level spacing distribution is a way to characterize these localization properties\cite{evangelou1,evangelou2,evangelou3,mirlin,graphene_crit,graphene_qm}.For example in 3D cubic lattices with disorder a critical disorder strength $W_c$ exists below which the spacing distribution follows the Wigner-Dyson distribution corresponding to chaotic wavefuctions with infinite localization length, forming the so called metallic phase\cite{evangelou2}. Above the critical point all states across the energy spectrum of the system become exponentially localized, the localization length becomes finite and the distribution of the spacing between successive energy levels reduces to the the Poisson form. 
This is an example where a universality is present, since the value of $W_c$ is determined only by the dimensionality of the system and is independent from the the microscopic details the system, for example how the disorder is implemented or the type of lattice used. For example the critical disorder strength is $W_c=16.5$ for 3D cubic lattices, while it reduces to $W_c=0$ for 2D square lattices or 1D chains, meaning that all states are exponentially localized even for infinitesimally small disorder for systems with dimensionality $D \leq 2$. The only exception is when the time-reversal symmetry is broken, 
for example at the presence of a magnetic field, or spin-orbit
coupling is taken into account. Both cases can result in 2D phases
with infinite localization length, akin to the metallic phases
of the 3D system\cite{evangelou1}.

In addition the spacing distribution reveals the quantum chaotic properties of a quantum system, based on the integrability of its classical counterpart\cite{berry1,berry2,stockman,courtney,nakamura,kottos1,kottos2,haake,smilansky}. For example classically non-integrable systems, like a quantum particle confined on a billiard of an irregular shape like a stadium or one with random boundary, follows the Wigner-Dyson spacing distribution for its energy levels. In contrast when the boundary is more regular, for example cyclic, the system becomes classically integrable and the energy level spacing of the quantum system follows the Poisson distribution. The Wigner-Dyson distribution signifies a quantum chaotic behavior and is one of the so called universality classes of random-matrix-theory (RMT), comprising of full random matrices with real elements representing the Hamiltonians of generic quantum systems with randomness, obeying time-reversal symmetry.

In order to investigate the localization and phase transition phenomena described above, for physical systems that contain dimensional fluctuations, we choose the most generic and simple type of random graphs, comprising of a fixed number of vertices n and edges m. The edges are randomly distributed among the vertices, resulting in $\binom{ \binom{n}{2}}{m}$ possible configurations(runs) of the graph. When all the configurations have an equal probability to appear $p=\frac{1}{\binom{ \binom{n}{2}}{m}}$, then we have the case of uniform random graphs $G(n,m)$, originally introduced by Paul Erdös and Alfred Rényi. We define the ratio of edges over vertices $R=\frac{m}{n}$ as a measure of  the spatial density of the graph. For $R < \frac{1}{2}$ the uniform graph consists of many small disconnected components. As more edges are added between its vertices there is a point at $R=0.5$ where a large dominating(giant) component appears, that contains most of the edges, and remains present for all ratios $R>\frac{1}{2}$\cite{frieze}. A spatial dimension emerges for this giant graph component, taking integer and fractional values, that fluctuate dynamically for the different graph configurations\cite{paper1,paper2}.

We examine how a quantum particle, for example an electron behaves as it propagates through the tight-binding lattice formed by the giant graph component that has a dynamical spatial dimension. This gives us the electronic properties of the graph. The Hamiltonian of the system can be written as
\begin{equation}
H = \sum_{<i,j>}^m(c_{i}^{\dagger}c_{j} +  h.c.)
\label{ham}
\end{equation}
where $c_{i}^{\dagger}(c_{i})$ is the creation(annihilation) operation for a particle at vertex i in the random graph. The indexes i,j are randomly sampled and create m pairs, representing the edges between the vertices in the graph. The resulting quantum system is a random tight-binding lattice with n sites and m hoppings randomly distributed between them, with the value of one. The matrix Eq. \ref{ham} is known also as the adjacency matrix of the graph. It is a random matrix with a fixed number of elements m (the number of edges), with the value of one, randomly distributed inside it. The randomness induces interference effects for the wavefunctions of a particle propagating through the graph, leading to localization phenomena as in the Anderson or RMT models.

\begin{figure}
\begin{center}
\includegraphics[width=0.9\columnwidth,clip=true]{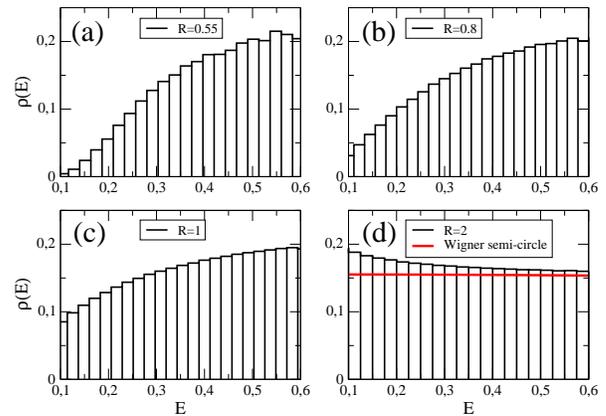}
\end{center}
\caption{The density of states(DOS) for different ratios R.
For small R in panels (a) and (b) the DOS resembles
the spectral properties of honeycomb/graphene lattices.
As R increases, and the graph Hamiltonians become denser,
the distribution approaches the Wigner semi-circle denoted
by the red line in panel (d), which is followed by also by
the real full random Hamiltonians of RMT.}
\label{fig1}
\end{figure}

In Fig. \ref{fig1} we show the density of states (DOS)
for different ratios R for a fixed graph size n=9000. For small R in Figs. \ref{fig1}a,b,c, corresponding to sparse graphs/Hamiltonians the DOS increases with the energy E, resembling for example the behavior
of honeycomb/graphene lattices\cite{paper1}. As R increases corresponding to denser graphs/Hamiltonians, the DOS approaches gradually the Wigner semi-circle \cite{paper1} denoted by the red line 
in Fig. \ref{fig1}d, which is also followed
by the DOS of full random Hamiltonians of RMT with real elements, belonging to the Gaussian-orthogonal-ensemble(GOE). This is reasonable since the Hamiltonians of dense graphs resemble the Hamiltonians of the GOE class of RMT.

\begin{figure}
\begin{center}
\includegraphics[width=0.9\columnwidth,clip=true]{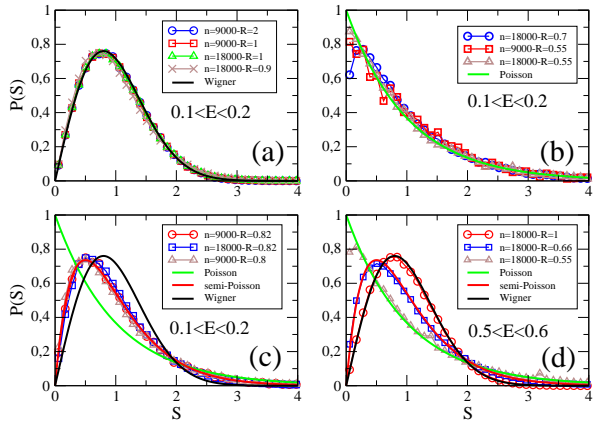}
\end{center}
\caption{The probability distribution of the spacing between successive
energy levels, after applying an unfolding technique for two different energy ranges. a) The Wigner-Dyson distribution is followed for large R, signifying a quantum chaotic behavior and a metallic phase with extended wavefunctions. The effect is independent from the graph size n. b)For small R, including cases near R=0.5
where the giant graph component appears, the spacing follows the Poisson distribution, signifying an integrable behavior with localized wavefunctions. c) In the intermediate regime between the metallic and the localized regime the spacing follows the semi-Poisson distribution. This is the critical regime where the phase transition between the metallic and localized regime occurs, as in 3D cubic lattices with disorder. The respective wavefunctions have a fractal structure. d) The chaotic, localized and critical behaviors manifest also in a different energy range.}
\label{fig2}
\end{figure}

In Fig. \ref{fig2}a,b,c we show the distribution of the spacing S between successive energy levels P(S), for different number of vertices n and ratios R, in the energy range $0.1 < E < 0.2$. The energy levels have been unfolded for 1000 configurations (runs) of the graph, ensuring that the spacing satisfies $\langle S \rangle=1$,
across the studied energy range. In Fig. \ref{fig2}a the spacing follows the Wigner-Dyson distribution, described by the formula
\begin{equation}
 P(s)=\frac{\pi}{2} s e^{-\frac{\pi}{4} s^2}  
\end{equation}
signifying a chaotic behavior as for ensembles of real full random Hamiltonians, which belong to the Gaussian-orthogonal-ensemble(GOE) class of RMT, obeying time-reversal symmetry. Similar chaotic behaviors have been observed for other types of graphs \cite{smilansky} and random Ising networks\cite{grabarits,pournaki}. In regular disordered systems, for example in 2D square lattices with on-site disorder, this chaotic behavior might appear also in the localized regime, when the linear length scale characterizing the system size is smaller than the localization length. In this case a gradual crossover from the Wigner
to the Poisson distribution is seen as the size of the system is increased. However in the graph  the Wigner distribution remains invariant under scaling as shown from the two cases n=9000 and n=18000 for ratio R=1, implying that the corresponding wavefunctions of the graph at the chaotic regime are extended, as in 3D cubic lattices for weak disorder below the critical point, where the localization length is infinite. 

As the graph becomes sparser by removing edges between its vertices the spacing distribution reduces to the Poisson form described by the formula
\begin{equation}
 P(s)=e^{- s} 
\end{equation}
as shown in Fig. \ref{fig2}b. This implies that the wavefuctions
have become localized and the graph resembles
integrable quantum systems. This behavior coincides also
with the appearance of the giant component in the graph
due to a structural phase transition at $R=0.5$. Below
this ratio the graph reduces to many small disconnected
components.

The semi-Poisson distribution
described by the formula,
\begin{equation}
 P(s)=4 s e^{-2 s} 
\end{equation}
is followed in the intermediate regime between the metallic 
and localized phases as shown in Fig. \ref{fig2}c. This is the critical regime where the phase transition occurs. In regular disordered system this critical regime is characterized by a fractal structure for the respective wavefunctions.

\begin{figure}
\begin{center}
\includegraphics[width=0.9\columnwidth,clip=true]{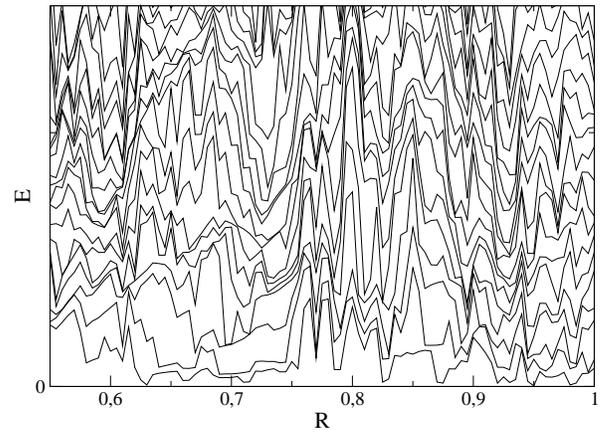}
\end{center}
\caption{The unfolded energy levels in the energy range $0.1<E<0.3$, for graph size n=9000 versus the ratio R. For large ratios near R=1 there is repulsion between the levels and very few crossing between them. The crossings become more frequent as R decreases and the system transitions from the chaotic to the localized regime
via a critical intermediate regime.}
\label{fig3}
\end{figure}

We have found chaotic, localized and critical
phases characterized by the Wigner and
Poisson, and semi-Poisson spacing distributions at various
energy ranges. The critical values of R 
where the transition between the two phases occurs,
depends on the energy of the system. 
For example by applying the unfolding technique
in the energy range $0.5 < E < 0.6$ we have found 
a phase transition from the Wigner to the Poisson distribution
at the critical ratio $R_c=0.66$ where the semi-Poisson distribution appears. The result is shown in Fig. \ref{fig2}d.

In Fig. \ref{fig3} we show the unfolded energy levels
versus the ratio R. For large R there is repulsion between the levels, while a few crossings between them start to appear as R decreases. The level crossings become more frequent as the system transitions gradually to the localized phase, characterized by the Poisson distribution for the level spacing.

\begin{figure}
\begin{center}
\includegraphics[width=0.9\columnwidth,clip=true]{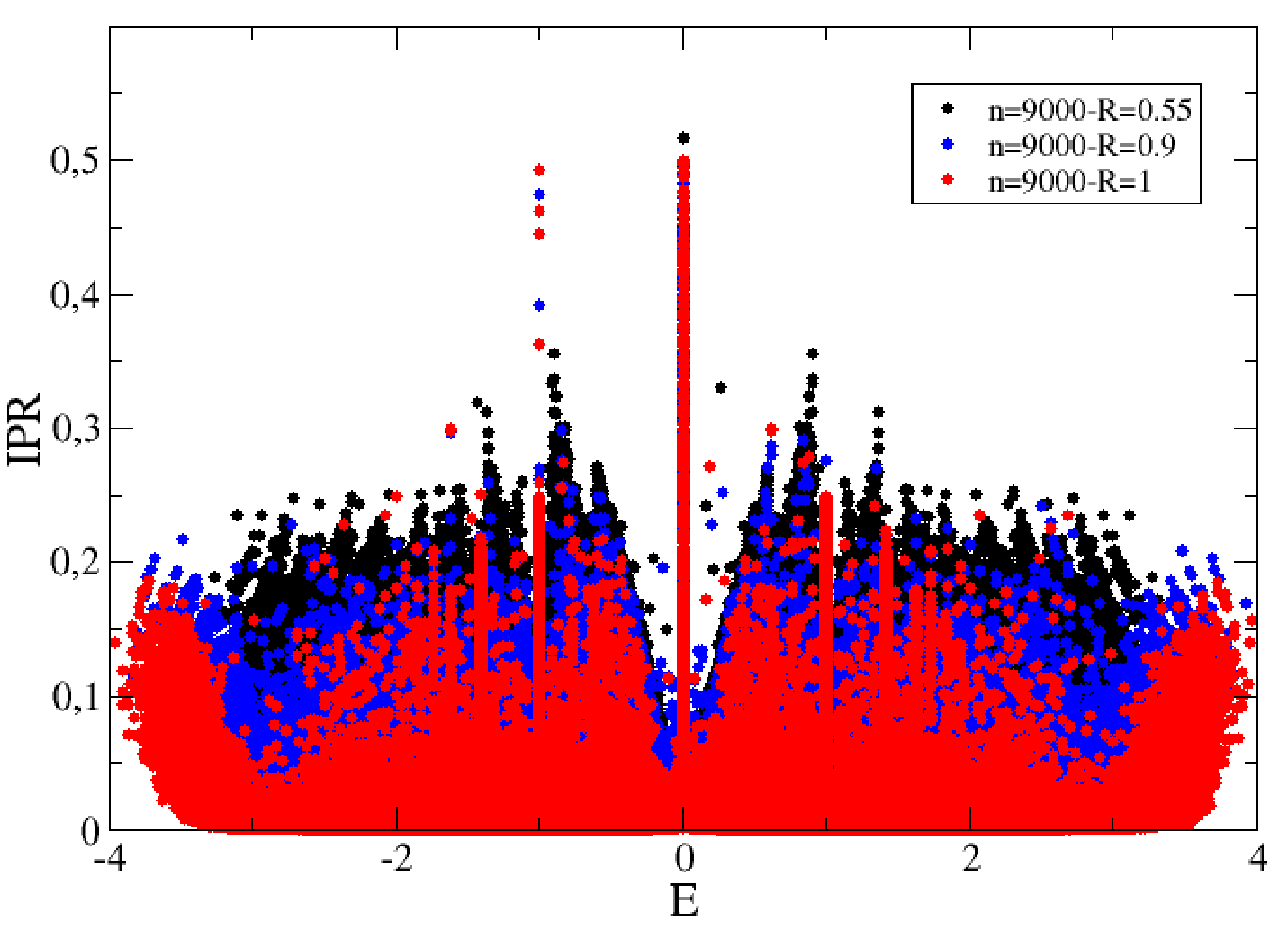}
\end{center}
\caption{The inverse-participatio-ratio(IPR) of the graph wavefunctions
versus the energy of the system E for 1000 runs. Smaller ratios R give larger values of IPR in average and therefore the corresponding wavefunctions become more localized as R decreases. The effect is less prominent at the edges of the energy spectrum.}
\label{fig4}
\end{figure}

The localization properties of the corresponding wavefunctions
can be estimated by calculating the inverse-participatio-ratio(IPR)
\begin{equation}
 IPR=\sum_i^n \left| \psi(i) \right| ^4  
\label{eq_ipr} 
\end{equation}
where $\psi(i)$ is the wavefunction amplitude at vertex i
of the graph while the sum runs over all the vertices.
There are two limits for the value of IPR that allows
us to use it in order to estimate the localization properties of the wavefunctions lying on the graph. Assuming that the wavefunction amplitude is uniformly distributed on the whole graph, then we have a normalized amplitude $\psi(i)=1/\sqrt n$ on each vertex and consequently Eq. \ref{eq_ipr} gives $IPR=1/n$, which becomes $IPR=0$ at the limit of large graph sizes $n \rightarrow \infty$. On the other hand when the wavefucntion is completely localized on one vertex, say j, with amplitude $\psi(j)=1$ ($\psi(i \neq j)=0$), we have $IPR=1$. Therefore extended wavefunctions give values of IPR near zero and localized wavefunctions give values of IPR near one. In Fig. \ref{fig4} we show the IPR for n=9000 for various m corresponding to different ratios, plotted along the whole energy spectrum for 1000 runs of the graph. The IPR decreases in average as R decreases, showing the onset to localization as the level spacing distribution transitions from the Wigner to the Poisson form.

A partial explanation of the transition to localization, 
and the relevant transition from chaotic to integrable level statistics, could be attributed to the decreasing average dimensionality of the graph,
as its spatial density decreases, by decreasing 
the ratio R. For example by following the scaling approach described
in Refs.\cite{paper1,paper2} to calculate the emergent spatial dimension D of the graph for n=9000, we have found average dimension$ \langle D \rangle=3.15$ for R=1 in the chaotic regime,  $ \langle D \rangle=2.64$ for R=0.82 at the critical regime, and
 $ \langle D \rangle=1.90$ for R=0.6 at the localized regime.
 Therefore the system transitions gradually from
 $D \simeq 3$ to $D \simeq 2$ as the spatial density of the graph reduces, which could provide a partial explanation of our result,
 since the majority of disordered systems with time reversal symmetry
 have all their states localized in two-dimensions(2D) while in three-dimensions(3D) there is a metallic regime with extended wavefunctions, below the critical point.

 In addition the average number of neighbors at each vertex, i.e. the average degree $\langle d \rangle$, which measures the average connectivity of the graph, is proportional to the ratio R as $\langle d \rangle=2m/n=2R$. The average connectivity is related to the topological spatial dimension of the graph. The wavefunctions are easier to localize for graphs with lower connectivity, since the electronic waves have fewer paths to travel at each vertex.
 
Also, the graph consists microscopically of many tree-like
structures and large cycles, which become more frequent as
the graph becomes sparser. This mechanism contributes also
to the localization properties. 

In conclusion we have shown how phase transitions
occur in uniform random graphs/networks, which have a non-fixed (fluctuating or dynamical) spatial dimension due to their inherent random geometry. We have concentrated on the quantum chaotic and localization properties, manifesting when the graphs are treated as random tight-binding lattices. The random geometry creates destructive interference effects leading to localization phenomena of the corresponding wavefunctions lying on the graphs, as in Anderson or RMT models. A transition from a metallic phase, with chaotic level statistics, to a localized phase, with integrable level statistics, is observed as the spatial density of the graph, characterized by its ratio of edges over vertices R, decreases. Since traditional phase transitions rely strongly on the spatial dimensionality of the system, our result shows that phase transitions with relevant universality behaviors can occur also in spatio-dimensionless physical systems
or ones with a fluctuating/dynamical spatial dimension. Furhter investigations of other types of random graphs with different types
of random geometries would be also interesting.

\end{document}